\documentclass{ringb99}
\usepackage{graphics}

\begin{document}
\title{A Radio Halo Found in the Hottest Known Cluster of Galaxies 1E0658-56}
\author{H. Liang}  
\institute{Physics Department, University of Bristol, Tyndall Avenue, Bristol, BS8 1TL, UK}
\maketitle

\begin{abstract}
We report the detection of a diffuse radio halo source in the hottest
known cluster of galaxies 1E0658-56 (RXJ0658-5557). The radio halo has
a morphology similar to the X-ray emission from the hot intracluster
medium.  The detection of such a strong radio halo in such a hot
cluster is further evidence to the link between X-ray temperature and
cluster-wide radio halos. We describe a new model for the origin of
cluster-wide radio halo sources involving a direct connection between the
X-ray emitting thermal particles and the radio emitting relativistic
particles.
 
\end{abstract}

\section{Introduction}
Diffuse cluster radio sources are found in a few X-ray luminous
clusters of galaxies. They are extended ($\sim 1$ Mpc), have low
surface brightness and steep spectra ($\alpha < -1$), which cannot be
identified with any one individual galaxy but rather with the whole
cluster. Diffuse cluster radio sources are generally separated into 2
classes: halos and relics, where halos are centred on the X-ray
emission (e.g. Coma-C is a proto-type radio halo source; Giovannini et
al. 1993) whereas relics are peripheral and exhibit stronger
polarisation than halos (e.g. A3667 has a large relic; R\"ottgering et
al. 1997). In this talk, we will concentrate on halos.  

Until recently, systematic surveys for radio halo sources have found
few examples and the total number of known halos was $\sim 5$ (Feretti
\& Giovannini 1996). Radio halo sources are thus considered to be rare
and owing to their small number, remain a poorly understood class of
radio sources even though the first example, Coma-C, was
discovered over 20 years ago. The spectra and some signs of
polarisation indicate that halo radio emission is dominantly by the
synchrotron process. However, the formation of radio halos 
remains a puzzle. The major questions involved are the origin of the
magnetic field and the relativistic particles, and why radio halos
occurs in some clusters and not in others. 

A number of models have been proposed to explain the formation of
radio halos (e.g. Jaffe 1977; Dennison 1980; Roland 1981). Most of
these early models suggest that ultra-relativistic electrons originate
either as relativistic electrons from cluster radio sources and
reaccelerated by in-situ Fermi processes or turbulent galactic wakes,
or as secondary electrons produced by the interaction between
relativistic protons (again from cluster radio galaxies) and thermal
protons.  However, the energetics involved are problematic and the
models could not always fit the observations. Harris et al. (1980)
first suggested that radio halos are formed in cluster mergers where
the merging process creates the shocks and turbulence necessary for
the magnetic field amplification and high energy particle acceleration. More
recently, Tribble (1993) showed that the energetics involved in a
merger is more than enough to power a radio halo. The halos thus
produced are transient which explains why they are rare.

In this talk, we will describe the properties of the radio halo found
in one of the hottest known clusters 1E0658-56, and suggest a new
model on the origins of cluster halos based on the radio and X-ray
properties of all confirmed halos.  Section 2 describes the radio halo
found in 1E0658-56; Section 3 discusses the model. Throughout the
talk we will use $H_{0}=50$ km\,s$^{-}$\,Mpc$^{-1}$, $q_{0}=0.5$ and
$\Lambda_{0}=0$.

\section{A Radio Halo in 1E0658-56}
Here we report the serendipitous detection of a strong radio halo
source in 1E0658-56.  The cluster was originally found in the Einstein
slew survey (Tucker et al. 1995), and subsequent optical observations
confirmed it to be a rich cluster at $z\sim 0.296$ (Tucker et
al. 1998).  It was selected as a candidate in the SEST campaign for
the detection of the Sunyaev-Zel'dovich effect (SZ effect; Andreani et
al. 1999). The cluster has high X-ray luminosity and
temperature. Recent ASCA results have shown it to be one of the hottest
clusters known (Tucker et al. 1998). The SEST observations show a
$\sim 4\sigma$ detection of the SZ effect at 1.2mm ($\sim$ 150 GHz)
and a $\sim 3\sigma$ detection at 2mm ($\sim$ 250 GHz) (Andreani et
al. 1999).  Subsequently, we obtained data on the ATCA at 8.8 GHz to
confirm the SEST detection of the SZ effect. The ATCA observations
were conducted using the special 210m array which proved to be an
excellent configuration for detecting low surface brightness diffuse
emissions. The 4.9 GHz and 8.8 GHz observations show  complex radio source structures, including at least two
extended, diffuse sources and an unusual strongly polarised ($\sim
60\%$ at 8.8 GHz) steep spectrum ($\alpha \sim -1.9$) source (Liang et
al. 1999). The diffuse radio sources make the detection of the SZ
effect difficult since the SZ effect is also extended and the
straightforward point source subtraction technique fails.  However,
since the radio sources in the cluster are interesting 
in their own right, we obtained further ATCA time to image the sources
at 1.3 and 2.4 GHz bands with a 6km and two 750m arrays.

\subsection{Radio Observations}
\begin{figure}
\resizebox{\hsize}{!}{\rotatebox{270}{\includegraphics{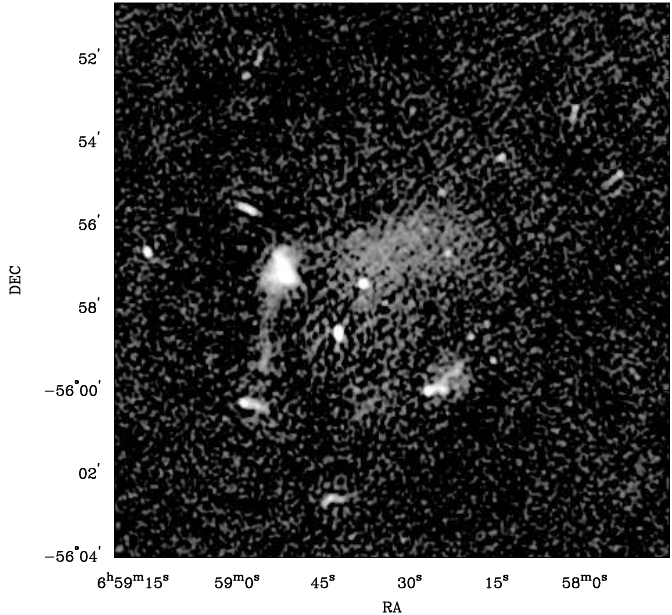}}}
\caption[]{A 1.3 GHz ATCA image towards the cluster 1E0658-56 at a resolution of $\sim 6^{''}$. The noise level in the image was  $\sim 44 \mu$Jy/beam. \label{21cm} }
\end{figure}

The cluster was observed at the ATCA at 1.3, 2.4, 4.9, 5.9 and 8.8 GHz
in several antenna configurations, so that similar uv coverage was obtained at all frequencies. The ATCA
has 5 antennas on a continuous rail-track over 3 km and a 6th antenna
6km away; it enables simultaneous observations in two frequency bands
each of bandwidth 128 MHz divided into 32 frequency channels. The halo
was clearly detected in all frequencies. The 2.4 GHz data were heavily
affected by interference and will not be considered further.  A 1.3 GHz
image of the cluster field is shown in Fig.~\ref{21cm}, where we see
the halo source at the cluster centre, a relic source to the east and
a number of tail sources on the periphery. A radio contour image at 1.3 GHz
of the halo, with embedded point sources subtracted, is shown
in Fig.~\ref{hrir}.
 
\begin{figure}
\resizebox{\hsize}{!}{\includegraphics{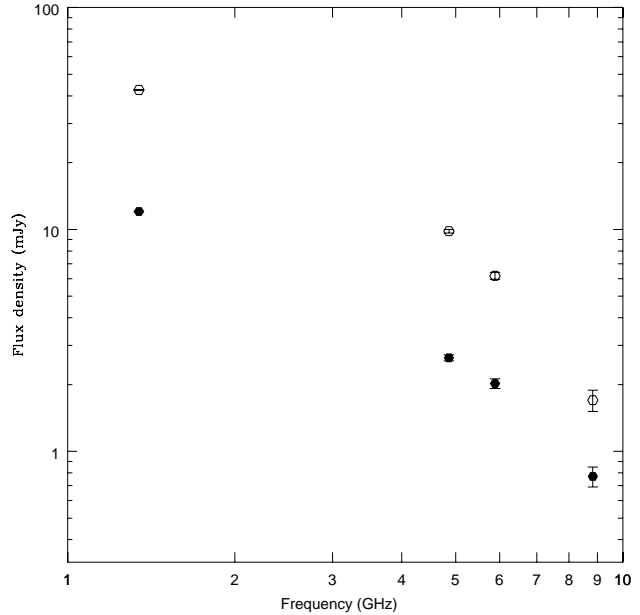}}
\caption[]{Radio spectra of the cluster halo in 1E0658-56. The filled
dots give the total flux within a central region  $\sim 210$
kpc in radius, whereas the open dots corresponds to the outer regions, between radii of $\sim 210$ kpc and $500$ kpc. The 8.8
GHz data points have not been corrected for the SZ effect.  \label{halosp} }
\end{figure}

To obtain the spectral indices of the halo, we took data from
approximately the inner uv-plane (baselines $< 3600\lambda$; $\lambda$
being the wavelength) for each available frequency. The observations
were planned such that roughly the same region in the uv-plane was
sampled at the various frequencies.  The data were tapered such that
the resultant beam size was $\sim 60^{''}$ at each frequency.  To form
a spectrum, the same area was used to integrate the flux at each
frequency. This area was selected to include only regions of obvious
emissions at all frequencies. The results are shown in
Fig.~\ref{halosp}.  The spectral index of the central part of the halo
is $\sim -1.18$ between 1.3 and 4.9 GHz, and $\sim -1.40$ between 4.9
and 5.9 GHz indicating possible spectral steepening at high
frequencies.  The data points in Fig.~\ref{halosp} have not been
corrected for the SZ effect. The SZ effect is expected to be strongest
at 8.8 GHz contributing $\sim -0.5$mJy in the central region, thus the
true spectra would be less steep at 8.8 GHz than what we see in the
plot.

\subsection{X-ray Properties}

The cluster was observed for 25 ksec by ASCA in May 1996.  Tucker et
al. (1998) analysed the ASCA GIS and SIS data as well as the ROSAT HRI
data and found the cluster to have a best fit temperature of
$kT_{x}\sim 17.4\pm 2.5$ keV and a bolometric luminosity of $L_{bol}
\sim (1.4\pm 0.3) \times 10^{46}$ ergs s$^{-1}$.

Since then, some PSPC data have become publicly available and it is
known that the SIS data below 1 keV suffered from inaccurate
calibrations, thus we re-estimate the cluster temperature by
simultaneously fitting the ASCA GIS and ROSAT PSPC data. The GIS and
PSPC data complement each other, since the GIS is more sensitive to the
cluster temperature and the PSPC is more sensitive to the soft X-ray
absorption.

We followed the standard ASCA procedure for screening the GIS2 and
GIS3 data as set out in ``The ABC guide to ASCA data reduction''. The
spectra were extracted from a circular region of radius $7.25'$
centred on the cluster, after the subtraction of a local background,
extracted from the same frame in areas where there are no obvious
emission. The spectra were regrouped to a minimum of 50 counts per
bin. The {\small XSPEC} package was used to fit a Raymond-Smith
spectrum (Raymond \& Smith 1977) with photoelectric absorption
(Morrison \& McCammon 1983) to the GIS spectra between 0.8 keV and 10
keV, leaving temperature, abundance, absorption (parametrised by N$_{H}$)
and the normalisation as free parameters. The best fit temperature was
$kT_{x} \sim 15.6^{+3.1}_{-2.3}$ keV.

We retrieved 4.7 ksec PSPC data observed in Feb. 1997 from the ROSAT
archive. A spectrum of the X-ray emission from the cluster gas was
extracted from the central $5^{'}$ radius after the subtraction of
discrete source and background contributions. The background was
estimated from an annulus centred on the cluster between a radius of
$8^{'}$ to $10^{'}$.

\begin{figure}
\resizebox{\hsize}{!}{\rotatebox{270}{\includegraphics*[115,44][556,726]{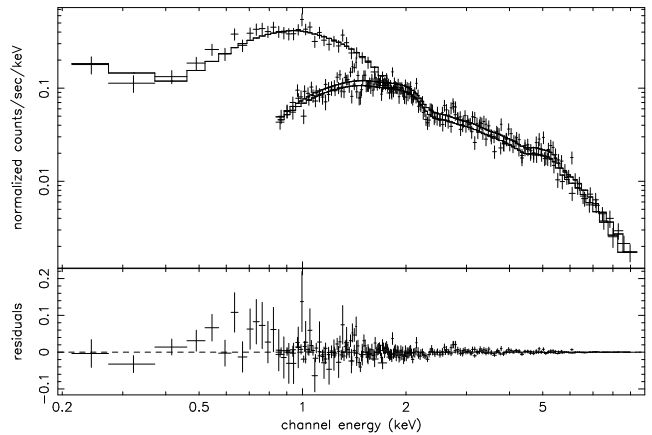}}}
\caption[]{X-ray spectra from ASCA GIS data and ROSAT PSPC data.  A
Raymond-Smith thermal spectrum with $kT_{x}\sim 14.5$~keV after a
convolution with instrumental responses is shown as a histogram on the
observed spectra (with error bars).  \label{ascasp}}
\end{figure}

A combined fit of a Raymond-Smith spectrum to the GIS and PSPC spectra
gave the best fit as follows: $kT_{x}=14.5^{+2.0}_{-1.7}$ keV,
$N_{H}=(4.1\pm0.5) \times 10^{20}$ cm$^{-2}$, abundance $= 0.48\pm0.24$
(see Fig.~\ref{ascasp}; errors correspond to 90\% confidence
limit). Thus it appears that the temperature of the cluster is lower
than estimated in Tucker et al. (1998), though it is still one of the
hottest known clusters.

\begin{figure}
\resizebox{\hsize}{!}{\includegraphics{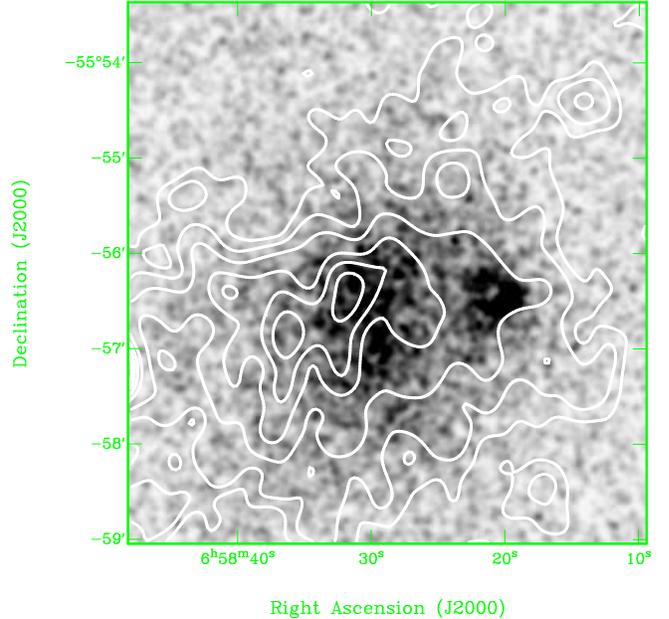}}
\caption[]{A contour plot of the radio halo at 1.3 GHz after the
subtraction of point sources  is overlaid on a
ROSAT HRI image smoothed with a Gaussian of $2^{''}$ width. The
resolution of the HRI is $\sim 5^{''}$. The radio image is smoothed to
a resolution of $\sim 20^{''}$. The contour levels are
0.2, 0.4, 0.8, 1.5, 2.0 \& 2.5 mJy/beam. The noise in the image is $\sim
80\mu$Jy/beam.
\label{hrir}}
\end{figure}

The cluster was also observed with the ROSAT HRI in 1995 for 58 ksec
(Tucker et al. 1998).  A HRI image of the cluster is shown in grey
scales in Fig.~\ref{hrir}.

\subsection{Discussions}
The radio halo source in 1E0658-56 is the strongest and largest known
at 1.4 GHz. Its flux density of $72.9\pm0.9$ mJy at 1.3 GHz
corresponds to a power of $(3.10\pm0.04)\times 10^{25}$\,W\,Hz$^{-1}$
within an area $\sim 1.2$\,Mpc$^{2}$. The largest linear extent of the
halo is $\sim 2$ Mpc. But this power is likely to be a lower limit
since the 1.3 GHz images show clear evidence of missing short
spacings: futher ATCA observations with the 210m configuration are
needed. The equipartition magnetic field was estimated to be
$B_{min}\sim 0.3\mu$G.

The HRI image of the cluster shows that it is a pre-merging system
with the two clumps clearly separated.  The radio halo has a similar
appearance and extent compared with the cluster X-ray emission (see
Fig.~\ref{hrir}), though less peaked. It is likely that shocks
produced through merging have provided the energy necessary for
accelerating the radiating electrons. However, the radio halo does not
resemble a shock but rather the distribution of thermal electrons that
produce the X-rays. 

\section{The Origin of Radio Halos}
\subsection{Are radio halos intrinsically rare?}
A number of surveys have been conducted to search for radio halos. The
earliest were conducted at Green Bank at 610 MHz (Jaffe and Rudnick
1979), at metre-waves 50-120 MHz (Cane {\em et al.} 1981) and at
Arecibo at 430 MHz (Hanisch 1982), but yielded few examples. Most of
the surveys selected either nearby Abell clusters (Hanisch 1982), or
clusters with known X-ray emission or radio sources. More recently,
Lacy {\em et al.} (1993) imaged a sample of radio sources from the 8C
38MHz survey (within $3.3^{\circ}$ of the North Ecliptic cap) using
the Cambridge Low Frequency Synthesis Telescope at 151 MHz but did not
find any new halo sources.

Recent X-ray
selected surveys of halos as well as observations aimed at detecting
the SZ effect have found many more halo candidates
suggesting that halos may not be as rare as they were thought to be.

Moffet and Birkinshaw (1989) first suggested that there may be a
correlation between the presence of SZ effect and radio halo sources,
since the only three clusters A2218, A665 and CL0016+16 which had SZ
effect detected at the time also had extended diffuse radio
emission. One of the strongest radio halos was found in A2163 in an
attempt to detect the SZ effect (Herbig and Birkinshaw 1994). Among
the 7 clusters observed at the ATCA for the SZ effect, 3 clusters show
clear evidence  of a radio halo (A2163, 1E0658-56 \&
S295), while another 2 clusters showed  faint extended
emission which can either be the result of the blending of faint discrete
radio sources or a faint halo (Liang 1995, Liang et
al. 1999). It is perhaps not surprising that searches for the SZ effect
have yielded more halo sources than the early surveys since candidates
for the SZ effect are hot X-ray luminous clusters. The SZ effect is also
cluster-wide, thus diffuse and extended like the halo sources. Any
observation designed to search for the SZ effect will optimise the
brightness sensitivity and thus favour the detection of halos.

Giovannini et al. (1999) in their correlation of NVSS images with the
catalogue of X-ray Brightest Abell clusters (XBAC; Ebeling et
al. 1996) found 29 candidate clusters with diffuse radio
emissions. They noticed a significant increase in the percentage of
diffuse radio sources in high luminosity clusters compared with  low luminosity clusters (27-44\% in clusters with L$_{x}>10^{45}$
ergs/s as compared with 6-9\% for L$_{x}<10^{45}$ ergs/s).

We conclude that radio halos are not intrinsically rare,
and appeared to be rare from the results of early surveys partly
because of the difficulty of detecting such low surface brightness
objects and partly because of the selection criteria. 

\subsection{The link between thermal and relativistic electrons}
\begin{figure}
\resizebox{\hsize}{!}{\includegraphics{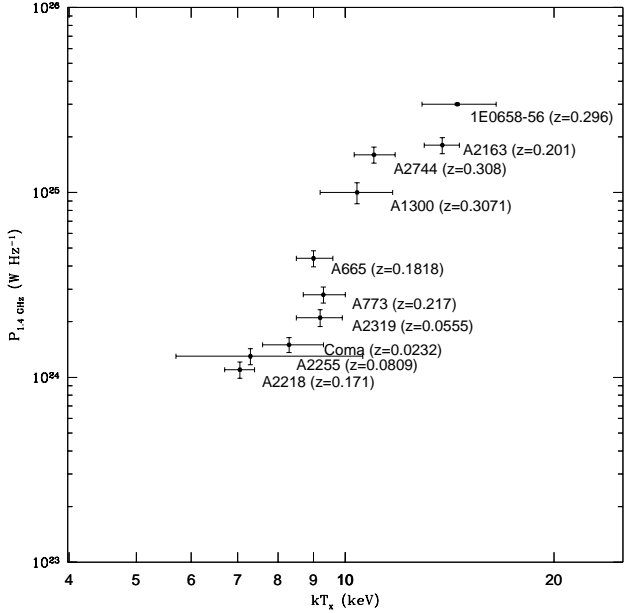}}
\caption[]{Radio power $P_{1.4}$ at 1.4 GHz versus the X-ray
temperature $kT_{x}$ for cluster-wide radio halo sources. The error
bars for the intracluster gas temperature correspond to a 90\%
confidence limit. Formal errors of $\sim 1\sigma$ are given for the
halo power where it is known, in cases where the errors are not
obviously stated in the literature, a 10\% error is assumed.  The data
are obtained from Reid et al. 1999, Pierre et al. 1999, Herbig \&
Birkinshaw 1999, Allen \& Fabian 1998, Mushotzky \& Scharf 1997,
Giovannini et al. 1999, Feretti et al 1997, Giovannini et al. 1993,
Markevitch et al. 1998, David et al. 1993 and Ferretti {\em et al.}
this proceedings.  The highest flux density is taken, where a halo has
been observed more than once at 1.4 GHz.
\label{halopt}}
\end{figure}

While Giovannini et al. (1999) found an increased occurrence of halos
in high X-ray luminosity clusters compared to the low luminosity ones,
they did not find a correlation between the radio power of diffuse
cluster radio sources and the cluster X-ray luminosity. Instead of
plotting radio power against X-ray luminosity, we examine the
relationship between halo radio power and cluster X-ray
temperature. Fig.~\ref{halopt} shows the 1.4 GHz integrated radio
power of cluster halos plotted against the cluster temperature,
demonstrating a steep correlation. Only well-confirmed radio halos
(not relic sources) are plotted, hence the sample of clusters shown is
by no means complete. The apparent rareness of halos can be explained
by the steepness of the relationship shown in Fig.~\ref{halopt}: only
clusters with a high X-ray temperature at moderate redshifts are
easily detectable. The surface brightness of halos goes as
$(1+z)^{-5}$ when taking account of the K-correction, thus the halo
surface brightness rapidly diminishes with increasing redshift. On the
other hand, halos with low redshift are also difficult to detect since
they tend to be resolved out in simple interferometric maps (or single
dish observations without a large beam-throw).

In the 3 well-imaged cluster halos (Coma, A2163 \&
1E0658-56), the extent and shape of the radio halo follows closely that of the
cluster X-ray emission (Fig.~\ref{hrir}; Deiss et al. 1997; Herbig and
Birkinshaw 1999).  Both the correlation shown in Fig.~\ref{halopt} and
the similarities between the radio and X-ray morphology indicate a
direct connection between the thermal particles and the relativistic
electrons responsible for the radio emission.

\subsection{A new model}
We propose a new model for radio halos where the thermal electrons in
the ICM provide the seed particles for acceleration to relativistic
and ultra-relativistic energies necessary for the production of
synchrotron emission. This differs from the existing theories
(e.g. Jaffe 1977) where the seed electrons diffused out of radio
galaxies. In the past, the possibility of the ultra-relativistic
electrons originating from thermal electrons accelerated to
relativistic energies has been dismissed, usually without much
justification.  We will examine this issue by looking at the arguments
put forward against the possibility of accelerating thermal electrons
to relativistic energies. One of the simple arguments was that such a
process would be present in every cluster and thus fail to explain the
perceived rarity of halos. As discussed earlier, such an argument is
rather simplistic: halos may be in every cluster and be detectable or
not according to their brightness, which appears to be related to the
temperature of the thermal gas.

A second argument against the ``thermal pool'' origin of the
ultra-relativistic electrons was that stochastic processes such as
Alfv\'en, turbulent and shock accelerations are only efficient in
accelerating electrons of energy at least several tens of keV
(e.g. Eilek \& Hughes 1991). This means that either the seed electrons
are mildly relativistic already, or an injection process is necessary
to create a substantial suprathermal tail in the electron energy
distribution. To circumvent the injection process (not yet
understood), it is attractive to invoke models that use
the already-relativistic electrons from radio galaxies as seed
particles to be re-accelerated. 

However, observationally, it has also been shown that while halo
candidates were found in nearly 30\% of the high X-ray luminosity
clusters in a survey for radio halo sources, none was found in the 19
clusters selected by their over-abundance of tailed radio sources
(Giovannini et al. this proceedings). This is further evidence that
the relativistic electrons are more likely to have originated from the
thermal pool of electrons than the radio galaxies: 
the appearance of halos is more closely linked to thermal X-ray
emission than the presence of tailed radio galaxies.

A closer examination of the argument for  an
injection process show that it may not always be necessary in a
cluster environment. According to Eilek \& Hughes (1991), electrons
need a minimum energy to be accelerated by Alfv\'en waves. In low
density environments, this minimum energy is already mildly
relativistic.  However, in a high density environment, where $n_{e} >
10^{-3} B^{2}_{\mu G}$\,cm$^{-3}$, an electron can resonate with
Alfv\'en waves at sub-relativistic energies of $\frac{1}{2} m_{p}
v_{A}^{2}/\cos^{2}\theta$ (where $v_{A}$ is the Alfv\'en speed and
$\theta$ is the pitch angle), i.e. energies of a few eV. The environments
of radio galaxies, including clusters, were considered to be of
low density  since the magnetic field strength $B$ was thought
to be a few $\mu$G, which means the density threshold is much higher than most cluster environments. Recent hard X-ray results from Beppo-SAX
and Rossi-RXTE for Coma and other clusters
have shown the magnetic field to be $B\sim 0.2\mu$G if the excess hard
X-ray emission is due to inverse Compton scattering of relativistic
electrons by the CMB (e.g. Fusco-Femiano et al. 1999; Rephaeli et
al. 1999; Valinia et al. 1999). Thus the density threshold is now
$\sim 4\times 10^{-5}$\,cm$^{-3}$, which
makes most parts of clusters high density environments. Therefore, it
is possible to accelerate thermal electrons in clusters through
resonance with Alfv\'en waves, and in the mean time, Dogiel
(this proceeding) has shown that it is possible to produce a
substantial suprathermal tail in the electron energy distribution
through second order Fermi acceleration in cluster environments.

Both merging activity and the thermal electron temperature maybe
responsible for the production of radio halos. A possible scenario
would be that the initial merging activity provides the energy for
accelerating electrons from the suprathermal tail of the energy
distribution (where the hotter clusters have more power) to
ultra-relativistic energies. After the shocks have disappeared, radio
halos like that of 1E0658-56 may be maintained by in-situ electron
acceleration in the residual turbulence.

Cooling flow clusters are thought to be relaxed and devoid of merging
activities. Most of the clusters shown in Fig.~\ref{halopt} are
non-cooling flow clusters.  This does not imply that mergers are the
{\it critical} element in radio halo formation, since it could be the
result of selection effects: cooling flow clusters are more
likely to have a significant central radio source than a non-cooling
flow cluster (e.g. Peres et al. 1998). To our knowledge, no cluster with a strong cooling flow  has been observed
properly to be able to test whether or not it follows the
$P_{1.4}-kT_{x}$ trend shown in Fig.~\ref{halopt}. To illustrate the
need for proper observations with high brightness sensitivities, we
give as an example, RXJ1347-11, a strong cooling flow with a very high
gas temperature of $\sim 12.5$\,keV (Allen and Fabian 1998) which has
been observed by the NVSS with no obvious detections. However, the
NVSS does not have enough brightness sensitivity to detect, in
RXJ1347-11, a halo similar to that in 1E0658-56 because of the high
redshift ($z\sim 0.45$) of the cluster and the high noise levels in
the NVSS image ($\sim 0.5$\,mJy per $45^{''}$ beam). 

\section{Conclusions}
We have found a strong radio halo source in the cluster 1E0658-56. At
a 1.3 GHz radio power of $(3.1\pm0.1)\times 10^{25}$\,W\,Hz$^{-1}$, it
is one of the most powerful radio halo sources. It has a steep
spectral index of $\alpha_{4864}^{1344} \sim -1.2$ typical of known
halos. The brightness distributions of the radio halo and X-ray
emission from the cluster gas are remarkably similar, suggesting a
direct relationship between the ultra-relativistic electrons
responsible for the synchrotron emission and the thermal intracluster
gas. As further evidence for the radio/X-ray connection, we have found
a steep correlation between the radio power of the halo and the X-ray
temperature of the intracluster gas ($P_{1.4}$ vs $kT_{x}$) from the
10 confirmed cluster radio halos.  We suggest a new explanation for
the origin of radio halo sources, where the high energy tail of the
thermal electron distribution is boosted to ultra-relativistic
energies, thus providing a natural link between the halo radio power
and X-ray gas temperature.  Finally, it is important for our
understanding of the origin of radio halo sources to establish the
robustness of the $P_{1.4}-kT_{x}$ correlation by observing a
temperature selected sample of clusters, and to test the mechanism by
searching for halos in clusters with strong cooling flows but high
temperature.

\begin{acknowledgements}
I would like to thank Sarah Maddison for help with the ATCA
observations. It is a pleasure to acknowledge the input from my
collaborators Mark Birkinshaw, Paola Andreani and Richard Hunstead, as
well as discussions with and encouragements from other conference
participants, especially R. D. Ekers, J. Eilek, V. Dogiel and A. Edge. The
Australia Telescope is funded by the Commonwealth of Australia for
operation as a National Facility managed by CSIRO.
\end{acknowledgements}

\end{document}